# Optimization of THz Microscopy Imaging

Katherine A. Niessen[a] and A.G. Markelz[ab]
[a]Department of Physics, University at Buffalo, SUNY, Buffalo, New York, USA
[b]Department of Structural Biology, University at Buffalo, SUNY, Buffalo, New York, USA
Email: kniessen@buffalo.edu

*Abstract*—**THz near field microscopy opens a new frontier in material science. High spatial resolution requires the detection crystal to have uniform and reproducible response. We present the THz near field spatial and temporal response of ZnTe and GaP and examine possible properties that give rise to the ZnTe degraded signal.**

## I. Introduction And Background

THz near field microscopy facilitates novel studies such as measuring the phase transition boundary for ferroelectrics and *in vitro* studies of mammalian cells[1,2], however, pixel to pixel response variation may not solely arise from the sample, but also from variation in electro-optical (EO) crystal response. To achieve highest spatial resolution of inhomogeneous systems requires the detection crystal to have uniform and reproducible response.

In Figure 1A we show an optical photo of a typical ZnTe crystal showing obvious defects throughout the crystal. These are micro bubbles from the crystal growth process. These bubbles are reported to have no effect on THz applications however it is unclear whether this applies to near-field detection[3]. To determine the optimal electro optic crystal for THz imaging we performed THz near field microscopy using a 500 μm thick (110) ZnTe crystal and a 300 μm thick (110) GaP crystal (both from MolTech GmbH) in the Planken geometry[4].

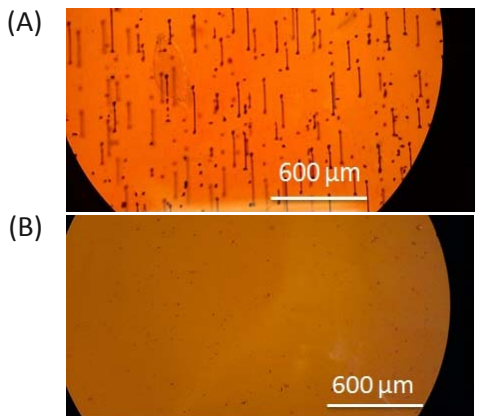

**Figure 1: (A)** Optical image of ZnTe crystal shows clear black striations at varying depths. **(B)** Optical image of GaP crystal shows no large visual defects.

The THz is generated using a GaAs photoconductive antenna illuminated with 800 nm, 100 fs pulsed light from a Ti:Sapphire laser at 80 MHz. Both EO detection crystals have a highly-reflective coating on the top surface and an anti-reflective coating on the bottom surface. The entire THz system is enclosed and the region is purged with nitrogen gas to remove water vapor. The power of the 800 nm at the detection crystal was kept constant (~5mW) for both crystals. The complete spectroscopy system is described elsewhere[5].

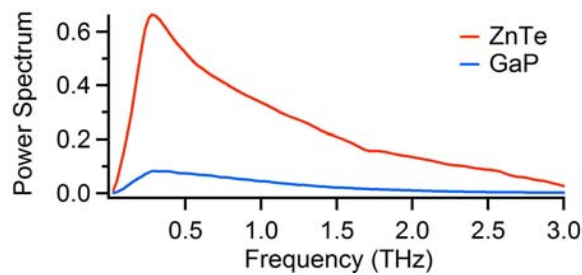

**Figure 2:** THz Power spectra, both measured at a single pixel of ZnTe and GaP, shows expected lower response for GaP.

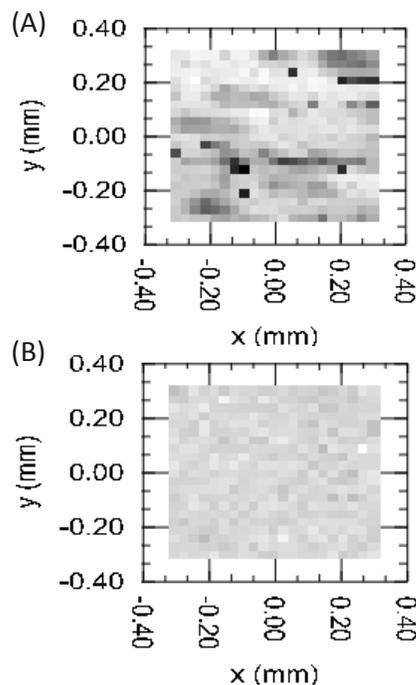

**Figure 3:** Normalized 2D image of the (A) ZnTe and (B) GaP crystal surfaces measured at 1THz.

## II. Results

In classifying the detection crystal surfaces we measured the signal variation across the surface and the change in signal over time. Figure 2 shows the power spectrum for a single pixel (30 μm x 30 μm) taken with the two crystals. The GaP has the expected lower response given its smaller EO coefficient. This could be compensated for by increasing the 800 nm power to the GaP crystal. The bandwidth is about 3 THz with both crystals. To measure the variation in the crystal surface, we performed a 2D raster scan in which the detection crystal moved relative to the THz and 800 nm beams. The THz power spectrum was obtained at each pixel. Figure 3 we shows a 2D scan at 1.0 THz using the (A) the ZnTe crystal and (B) the GaP crystal. The color scale is greyscale with signal normalized to the image maximum as white and zero signal as black. As seen in the figure the pixel to pixel response can vary as much as 98% over the 600 μm x 600 μm region for the ZnTe, whereas this variation is only 26% for the GaP. Furthermore we see in Figure 4 the time dependence for the signal at a single pixel is considerably stronger for the ZnTe than the GaP. The specific cause of the decrease in signal over time has not yet been determined. We will discuss the role of crystal stability in optimizing THz microscopy imaging systems based on these results.

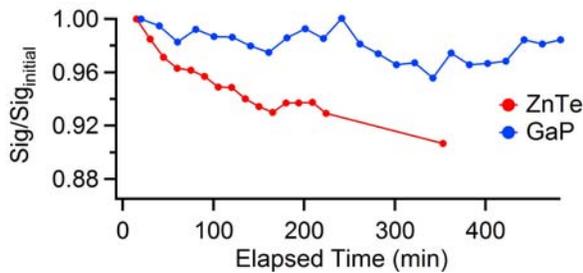

**Figure 4: Change in peak signal over time for ZnTe and GaP, normalized to first measurement. Crystal was constantly under 800 nm and THz radiation**.

### III. DISCUSSION

The higher nonuniformity and larger decrease in signal over time of the ZnTe crystal could be due its lower damage threshold than GaP. The 800 nm light with a 3mm beamwidth is focused on the ZnTe with a microscope objective with focal length 8.1mm. This gives a spot size of 2.8 μm. With this spot size, the calculated power density is about 10 GW/cm$^2$. The actual power density at the EO crystal is expected to be less do to diffraction effects since the focus is within the crystal and not at the surface. The overall resolution of the system is 10-30 μmIt is widly accepted that ZnTe has a much lower damage threshold than GaP but there is much disagreement with the measured values. The damage threshold of ZnTe, varies from 1.0 to 1.6 MW/cm$^2$ with 1.06 μm light[6] to 100 GW/cm$^2$ [7]. Damage, described as a black spot on the crystal surface, at less than 100 GW/cm$^2$ has also been seen after continuous illumination[8]. GaP has a measured damage threshold of 675 MW/cm$^2$ at 1.06 μm [9] to 70 GW/cm$^2$ measured with 1.04 μm[10]. We do not see any additional defects on the crystal surfaces after illumination. Photo-induced precipitation of tellurium in ZnTe crystals around 1 kW/cm$^2$ has been seen with CW 647 nm light[11]. No known measurements have been done to determine if crystalline Te is present in ZnTe crystals used in THz applications, which could affect the EO properties of the crystal.

Continuous illumination could be causing crystal heating, which may cause a decrease in detected signal. With the spot size of our system, heating of the ZnTe would be low according to Lin et al.[8]. However, the black striations in the ZnTe may have a higher absorption of 800 nm light, which could cause more heating damage. Our system does not allow for us to determine where the black striations are relative to the focus during a THz measurement.

### IV. CONCLUSION

Compared to the GaP crystal, we see a large nonuniformity in a 2D image of ZnTe at 1THz as well as a larger time dependence of the measured peak signal. The GaP crystal has a higher uniformity and stability, but has a lower amplitude power spectrum. This tradeoff between crystal uniformity and detected signal should be considered when choosing a crystal for a specific measurement, i.e. imaging or single pixel measurements.

ACKNOWLEDGMENT

We thank the National Science Foundation MRI^2 grant DBI2959989 for support.